\documentclass[aps,prd,twocolumn,amsfonts,amssymb,amsmath,showpacs,letterpaper,superscriptaddress,nofootinbib,altaffilletter,fontenc]{revtex4-1}

\usepackage[citecolor=blue,linkcolor=blue,
colorlinks=true,urlcolor=blue]{hyperref}
\usepackage{graphicx}
\usepackage{amssymb}
\usepackage{amsmath}
\usepackage{longtable}
\usepackage{rotating}
\usepackage{color}
\usepackage{epsf}
\usepackage{fancyhdr}
\usepackage{ifthen}
\usepackage{slashed}
\usepackage{xspace}
\usepackage{tabularx}
\usepackage{url}
\usepackage{hyperref}
\def\be{\begin{equation}}
\def\ee{\end{equation}}
\def\bi{\begin{itemize}} 
\def\ei{\end{itemize}}
\def\ben{\begin{enumerate}}
\def\een{\end{enumerate}}

\begin{document}

\title{Wavescan: Multiresolution Time-Frequency Transform of Gravitational-Wave Data}

\author{S.~Klimenko}

\affiliation{University of Florida, P.O.Box 118440, Gainesville, Florida, 32611, USA}

\begin{abstract}

Identifying transient signals embedded in non-stationary noise requires analyzing the time-dependent spectral components of the observed time series. The time-frequency distribution of the signal power can be estimated with Gabor atoms, or wavelets, localized in time and frequency by a window function. Such analysis is constrained by the Heisenberg-Gabor uncertainty principle, which limits the simultaneous time and frequency localization achievable with individual wavelets. Moreover, the resulting time-frequency distribution is subject to temporal and spectral leakage, limiting the identification of sharp or rapidly varying features in the power spectrum. This paper introduces a time-frequency transform that uses a stack of wavelets to scan local power across multiple scales. At each time-frequency location, a wavelet least affected by the leakage is selected from the stack to obtain high-resolution localization of power. The resulting wavelet scan ("wavescan") extends conventional multiresolution analysis by enhancing time-frequency localization and mitigating local power distortions caused by temporal and spectral leakage. The paper describes the principal components of the wavescan framework, including the estimation of the time-varying spectrum, identification of transient signals in the time-frequency data, and reconstruction of the corresponding time-domain waveforms. To demonstrate the performance of the method, the wavescan transform is applied to gravitational-wave data from the LIGO detectors.

\end{abstract}


\maketitle

Analysis of time series with time-dependent spectral density is a challenging problem in signal processing. Often signals present in noisy data may have rapidly varying temporal and spectral components with a sharp localization of the signal power both  in time and frequency. To obtain the time-frequency (TF) representation of the time series ${x}(t)$, the most widely-used is the short-time (window) Fourier transform (STFT)~\cite{AllenRabiner1977}
\begin{equation}
\label{stft}
   X(\tau, f) = \int_{-\infty}^{\infty} {x}(t) w(t-\tau) e^{- i2\pi{ft}} \; dt, 
\end{equation}
where the analysis window $w(t-\tau)$ of a finite duration captures the signal power centred around time $\tau$ and frequency $f$. The time-frequency distribution (or spectrogram) is estimated as $P(\tau,f)=|X(\tau,f)|^2$. The window function $w$ determines the STFT root-mean-square resolutions in time $\delta{t}$ and frequency $\delta{f}$ constrained by the Heisenberg–Gabor uncertainty 
$2\pi\delta{t}\delta{f}\geq{1/2}$~\cite{Gabor1946}.
In general, it is not possible to optimally describe all temporal and spectral components in the data by selecting a single window. 
For example, a gravitational wave (GW) signal from two colliding black holes~\cite{GW150914properties} has a chirp waveform, with amplitude and frequency increasing with time until the black holes merge. To track such signals, the transform requires the time resolution of $\delta{t}\ge{1}$~s 
at the early stage of the binary evolution and 
$\delta{t}\le{0.01}$~s at the end of the binary evolution when the system merges.
For chirping GW signals, this limitation can be partially alleviated by using  the conventional wavelet transforms~\cite{Mallat2008}, where the wavelet basis is obtained by scaling a single  mother wavelet and all wavelets have the same number of cycles. As a result, the wavelet duration decreases  as its frequency increases, improving the estimation of the GW power at the merger stage. Such wavelets are widely used for the time-frequency visualization of detected GW signals~\cite{Qscan2004, ROBINET2020100620}. 
However, they require individual tuning of the transform parameters for each GW signal, and for other transient signals, the conventional wavelet representation could be suboptimal. 

To overcome the trade-off between the time and frequency resolution and improve the time-frequency localization of 
signal power, several high-resolution techniques have been proposed~\cite{Cohen1989}, starting with the Wigner-Ville (WV) transform~\cite{Wigner1932, Ville1948}. The WV transform provides one of the most accurate time-frequency representations, but it produces the interference (cross-term) artifacts for multi-component signals.
These artifacts can be partially suppressed by applying smoothing kernels, which leads to bilinear time–frequency distributions $P(\tau,f)$ belonging to the Cohen's class~\cite{Cohen1966}. For appropriate kernels, these distributions preserve the marginals
\begin{equation}
\label{marg}
   |x(t)|^2 = \int{P(t,f)df} \;, ~~|S(f)|^2 = \int{P(t,f)dt} \;, 
\end{equation}
where $S(f)$ is the Fourier transform of ${x}(t)$. A different approach to the high-resolution techniques is based on the multiple STFT spectrograms obtained with different time-frequency resolutions that are combined together into a single spectrogram~\cite{MulSpect1994,superres,GeomMean,superlets}. Such spectrograms improve the visualization of the time-frequency data, but still, they are affected by the artifacts due to the temporal and spectral leakage, and usually, they do not satisfy the marginals.

The techniques discussed above seek to construct a joint time-frequency distribution (TFD) that provides a physically meaningful representation of a time-varying spectrum. TFDs are valuable for identifying and characterizing transient signals, particularly when their waveforms are not known a priori and matched-filtering methods cannot be readily applied. In practice, the choice of TFD is often application-dependent because different signal morphologies and noise environments require different compromises among time-frequency resolution, interference suppression, and statistical robustness. 
For example, to identify and extract a GW signal~\cite{Klimenko2016}, the time series from several detectors must first be mapped into the time-frequency domain with a transform that preserves the polarization state and phase structure of the signal. This multi-dimensional data is then filtered to extract the GW signal, from which the time-domain waveforms are reconstructed with the inverse transform.

Such model-agnostic analysis, filtering, and synthesis of gravitational waves had been implemented in the GW data analysis pipeline called coherent WaveBurst~\cite{Klimenko2016, Klimenko2008,Klimenko2005}. For the TFD estimation, it uses the orthonormal Wilson-Daubechies-Meyer (WDM) transform~\cite{Necula2012}, where the basis functions are constructed to have the same resolution across the frequency scale. Similar to the STFT and the traditional wavelets, the WDM transform at a fixed resolution is not capable of efficiently capturing a multi-component signal. To improve the identification of such signals, the WDM transform is performed at several (typically 6-8) resolutions, producing an oversampled time-frequency representation of data. The WDM amplitudes from  all resolutions are analyzed together to identify and reconstruct GW signals in the WDM domain~\cite{Klimenko2016}. The time-domain GW waveforms can also be reconstructed by using the inverse WDM transform.

The WDM technique had been successfully used for the first direct observation of gravitational waves from the merger of two black holes~\cite{GW150914}, and for detection of gravitational wave signals~\cite{GWTC1,GWTC2,GWTC3} in the data collected by the LIGO and Virgo detectors~\cite{AdLIGO2015, AdVirgo2014}.
However, despite its multiresolution approach, the WDM transform is affected by the temporal and spectral leakage hindering the efficient identification of weak GW signals in the time-frequency data. 

The goal of this work is to obtain the leakage-free time-frequency distribution. It is based on the multiple STFTs where the optimal resolution at each time-frequency location is dynamically selected according to the wavescan algorithm. The resulting wavescan transform can be used for the model-agnostic analysis, filtering, and synthesis of gravitational-wave signals. 
    
The rest of the paper is organized as follows. Section~\ref{wavescan}
introduces the wavescan method, including the description of the wavescan transform and the definition of the high-resolution time-frequency distribution. Section~\ref{cluster} introduces the excess power and the cross-power statistics used to identify signals in the time-frequency data. Section~\ref{waveform} discusses the reconstruction of the time-domain signals from the time-frequency data, e.g. the inverse wavescan transform, followed by the conclusion.

\section{Wavescan}
\label{wavescan}

The wavescan method applies multiple STFTs to discrete data using Gabor atoms with analysis windows of different durations. These localized atomic waveforms are referred to throughout this paper as wavelets. For a given time-frequency location $(t_n,f_m)$, and the analysis window $w_l$, there are two wavelet amplitudes $a_{lnm}$ and $\tilde{a}_{lnm}$ given by the inner products
\begin{equation}
\label{inner}
a_{lnm}=\langle{{x}},\psi_{lnm}\rangle\;, 
\tilde{a}_{lnm}=\langle{{x}},\tilde{\psi}_{lnm}\rangle\;, 
\end{equation}
of ${x}(t)$ with the symmetric $\psi_{lnm}$ and antisymmetric $\tilde{\psi}_{lnm}$ wavelets normalized to unit norm
\begin{eqnarray}
\label{FGT1}
\psi_{lnm}(t) \propto w_l(t-t_n)~cos[\omega_m(t-t_n)] \;, \\ 
\label{FGT2}
\tilde{\psi}_{lnm}(t) \propto w_l(t-t_n)~sin[\omega_m(t-t_n)] \;.
\end{eqnarray}

The wavelets are organized in a stack where the atoms with different resolution $l$ are centered at the same time $t_n$ and frequency $f_m$. A wavescan stack is constructed by scaling the duration of the window function $w_l$: $T_w[l]={\alpha}^{l-1}T_o$, $1\leq{l}\leq{L}$,  where $T_o$ is the duration of the shortest window, $L$ is the total number of resolutions in the stack, and $1 < \alpha \le 2$ is the scaling parameter. 
The window can be selected to have the Gaussian shape, which is optimally localized with $2\pi\delta{t}\delta{f}={1/2}$. However, the Gaussian window is not localized in time, and in general, other close to optimal window functions with compact support can be considered. Specifically, in this study, the Blackman-Harris window is used.

The wavescan stacks form a regular sampling lattice 
\begin{equation}
\label{lattice}
  t_n=n\Delta{t},~~~f_m=m\Delta{f} \;,   
\end{equation}
where $\Delta{t}$ and $\Delta{f}$ are the sampling steps in time and frequency. The wavescan lattice is highly oversampled with $\Delta{t}<{\delta{t}_{min}}$ and $\Delta{f}<{\delta{f}_{min}}$, where  $\delta{t}_{min}$ is the effective duration of the shortest wavelet and $\delta{f}_{min}$ is the effective bandwidth of the longest wavelet in the stack. 
The power $P_{lnm}$ and amplitude $A_{lnm}$ of the wavescan data are defined as
\begin{eqnarray}
\label{TFDnu}
P_{lnm}=\frac{1}{2}\left[ a^2_{lnm}+\tilde{a}^2_{lnm}\right],
~~A_{lnm}=\sqrt{P_{lnm}} \;. 
\end{eqnarray}
\noindent

\begin{figure*}[t]
    \includegraphics[width=0.49\textwidth]{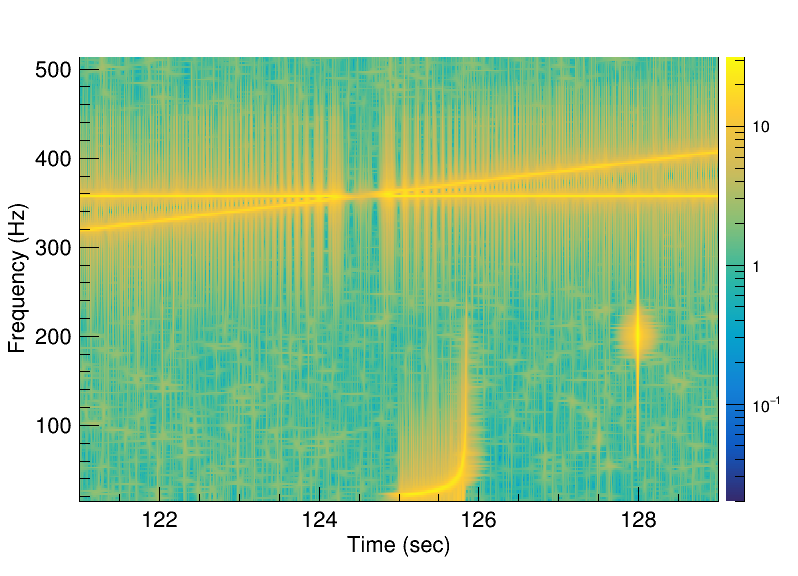}
    \includegraphics[width=0.49\textwidth]{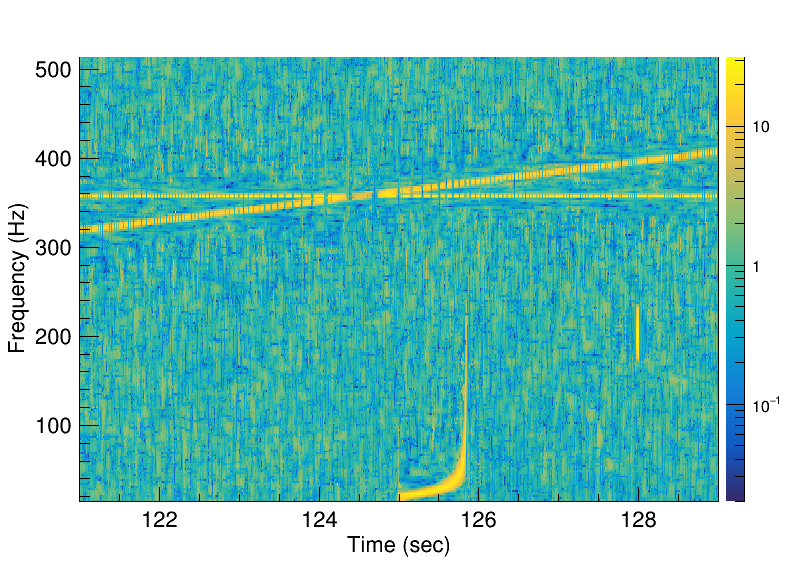}
    \caption{Time-frequency distributions of test data: left - maximum power $A_{max}$, right - wavescan.  }
   \label{fig1}
\end{figure*}

To illustrate the issue with the temporal and spectral leakage, 
the following test data are used, including the four different simulated signals: a) a monochromatic line, b) a linear chirp, c) a broadband cos-Gaussian burst at $t=128$ seconds, with the duration of 0.01 seconds, and d) a GW chirp at $t=125-126$ seconds. All these signals are added to white Gaussian noise with the unity variance. The parameters of the wavescan stack are: $T_o=1/64$~s, $L=6$, $\alpha=2$, $\Delta{t}=1/256$~s, $\Delta{f}=4$~Hz.  The left panel of Figure~\ref{fig1} shows the TFD estimator $A_{max}(t_n,f_m)= \max_{l}(A_{lnm})$, which may look like a natural choice for identification of weak signals in the noisy data. However, the $A_{max}$ TFD exhibits a problem: both the deterministic signals and the noise show severe artifacts due to the temporal and spectral leakage.

In wavescan, these TFD artifacts are mitigated by selecting, at each stack location ($n,m$) omitted below for brevity, a wavelet that is less affected by the leakage. The procedure for identifying the least affected wavelet differs for stochastic noise and deterministic signals. 

We first consider the case of stochastic noise. Exploiting the overlap between wavelets
\begin{equation}
\label{overlap}
O_{ij}={\langle\psi_{inm},\psi_{jnm}\rangle}\;,
~~\tilde{O}_{ij}={\langle\tilde{\psi}_{inm},\tilde{\psi}_{jnm}\rangle}\;. 
 \end{equation}
 the cross-talk coefficient between the wavelet $i$ and all other wavelets in the stack is defined as follows
\begin{equation}
\label{xtalk}
c_{i}=\frac{1}{\sum_{j\neq{i}}{\!{A_iA_j}}} \sum_{j\neq{i}}\left( a_i{a_j}O_{ij}\!+\!\tilde{a}_i\tilde{a}_j\tilde{O}_{ij}\right)\!. 
\end{equation}
The cross-talk statistics for wavelet selection are defined  as
\begin{equation}
\label{xtalk2}
\lambda_c = c_{\mathrm{max}}+c_{\mathrm{min}} \;,~~~~\lambda_p = c_{\mathrm{Pmax}}+c_{\mathrm{Pmin}} \;,
\end{equation}
where $c_{\mathrm{max}}$ and $c_{\mathrm{min}}$ are the maximum and the minimum values of $c_i$ in the stack, and $c_{\mathrm{Pmax}}$ and $c_{\mathrm{Pmin}}$ are the cross-talk values  corresponding to the maximum and the minimum power in the stack respectively.
The statistic $\lambda_c$ distinguishes local dips from local bumps in the noise TFD. For Gaussian noise, its distribution is bimodal, with an optimal threshold of $\lambda_c = 0.5$ separating the two peaks. Values below this threshold indicate a local dip, prompting the selection of the wavelet with cross-talk $c_{\mathrm{min}}$. Values at or above the threshold indicate a local bump, prompting the selection of the wavelet with cross-talk $c_{\mathrm{max}}$. This selection rule suppresses temporal and spectral leakage in the TFD of stochastic noise but is less effective for deterministic signals.

Because of the strong overlap between adjacent wavelets within a stack, the estimated power is expected to vary smoothly across resolutions unless the stack is affected by leakage. This behavior can be quantified by the maximum power drop in the stack, defined as
\begin{equation}
\label{powf}
\delta_p=\max_{l}\left\{{\frac{|P_{l}-P_{l+1}|}{MIN(P_{l},P_{l+1})+\sigma^2}}+\frac{1}{2}\right\} \;,
\end{equation}
where the variance of the noise $\sigma^2$ is used to suppress the large variations of $\delta_p$ due to the noise fluctuations. Without loss of generality, we consider whitened data with unit noise variance, $\sigma^2=1.$

\begin{figure*}
   \includegraphics[width=0.49\textwidth]{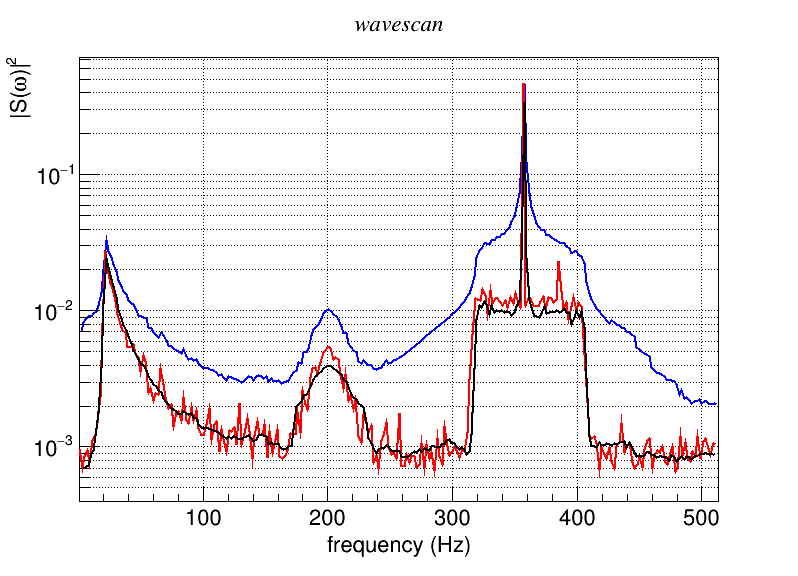}
   \includegraphics[width=0.49\textwidth]{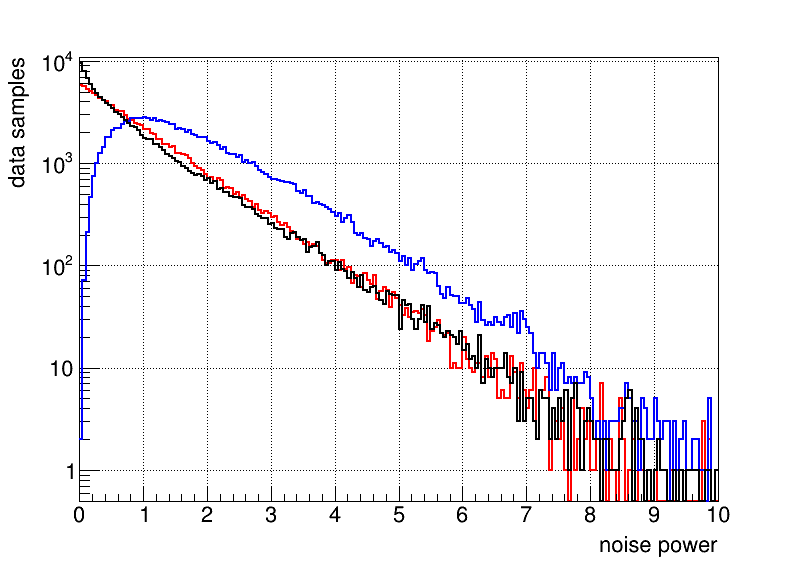}
   \caption{Left panel: Power spectral density distribution $|S(\omega)|^2=\int{P(t,\omega)dt}$ of the test data for the wavescan(black), maximum-power $A_{max}$ (blue) and WDM(red) estimators for the frequency binning of 2 Hz. Right panel: Comparison of the noise power distributions from the wavescan (black), maximum-power $A_{max}$ (blue),  and Gaussian noise with unity variance (red).}
   \label{fig2}
\end{figure*}

Wavelet stacks exhibiting a large power drop $\delta_p>2$, or satisfying $\lambda_c-\lambda_p<0$, are considered to be affected by leakage. If only one of these conditions is met, the wavelet with the smaller power is selected from the two wavelets with cross-talk values $c_{\mathrm{min}}$ and $c_{\mathrm{max}}$. If both conditions are met, the leakage is more severe, and the wavelet with the minimum power in the entire stack is selected. The threshold on $\lambda_c-\lambda_p$ is strictly defined to isolate the subset of data affected by leakage. The threshold on $\delta_p$ is chosen to exclude most noise samples from this selection and preserve the exponential tail of the noise distribution (see Figure~\ref{fig2}). These criteria suppress most leakage artifacts induced by deterministic signals.

In addition, when a stack is not affected by leakage, the wavelet amplitudes are expected to evolve in a predictable manner: 
\begin{equation}
\label{pow}
a_l\sim{a_I O_{Il}} \;,~~~~~ \tilde{a}_l\sim{\tilde{a}_I \tilde{O}_{Il}} \;, 
\end{equation}
where ($a_I,\tilde{a}_I$) are the wavescan amplitudes corresponding to the maximum power in the stack, and ($O_{Il},\tilde{O}_{Il}$) are the overlap coefficients between wavelets $l$ and $I$. Such wavelet stacks are therefore characterized by a small power imbalance
\begin{equation}
\label{powD}
\Delta_p=\frac{\sqrt{2}}{a^2_I+\tilde{a}^2_I}~{\sum_l{[(a_IO_{Il}-a_l)^2+(\tilde{a}_I \tilde{O}_{Il}-\tilde{a}_l)^2+2\sigma^2}]} \;.
\end{equation}
When a wavelet stack satisfies $\delta_p\Delta_p<1$, the wavelet with the maximum power is selected. This condition recovers the optimal resolution for deterministic signals that might be missed by the other criteria, without affecting wavelet selection for stochastic noise. In the adopted wavescan algorithm, the threshold on $\delta_p\Delta_p$ allows little flexibility. Lowering it introduces discontinuities within regions of high power, whereas raising it begins to admit pixels affected by leakage.
Figure~\ref{fig1}(right) shows the TFD obtained with the wavescan transform of the test data.  Compared to the maximum-power estimator $A_{max}$, shown in the left panel of Figure~\ref{fig1}, the wavescan transform provides an improved signal localization and substantially reduces leakage-induced artifacts in the time–frequency distribution. 
This is confirmed by the frequency marginals, $\int{P(\tau,f)dt}$, shown in Figure~\ref{fig2}(left), which are compared against the reference power spectral density obtained from the WDM transform. The wavescan and WDM distributions agree well, whereas the $A_{max}$ TFD is affected by leakage artifacts, causing its frequency marginal to significantly overestimate the power spectral density. Figure~\ref{fig2}(right) shows that the wavescan mostly preserves the exponential power distribution expected for Gaussian white noise, while the $A_{max}$ TFD estimator has a distorted noise distribution. 
\begin{figure*}[t]
   \includegraphics[width=0.49\textwidth]{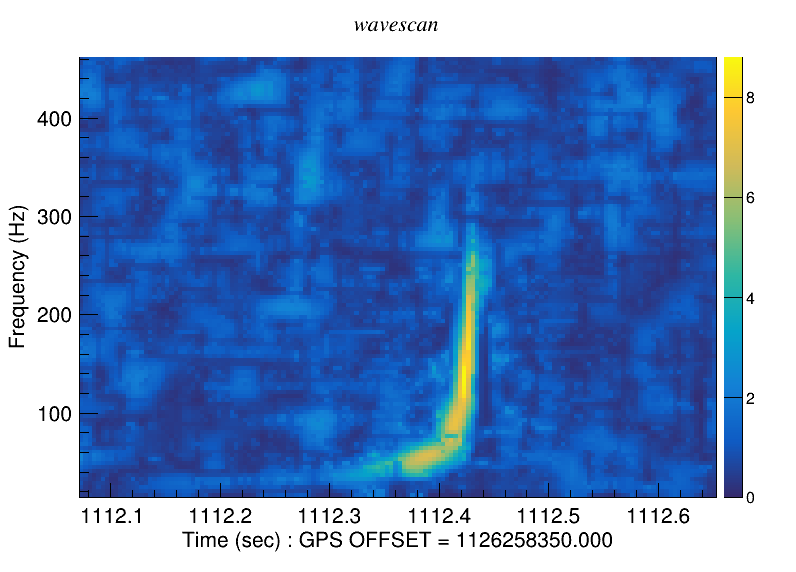}
   \includegraphics[width=0.49\textwidth]{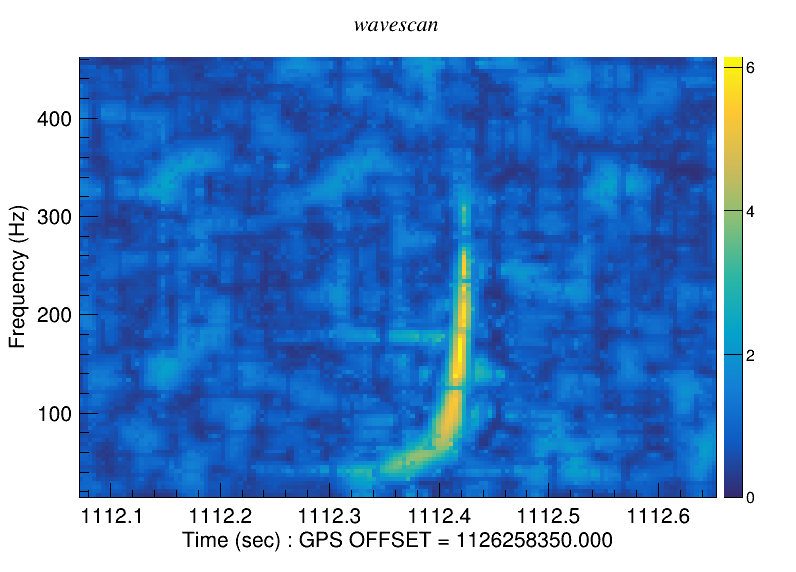}
   \caption{Wavescan spectrograms for the first GW signal GW150914 detected by LIGO~\cite{GW150914}: left-Hanford detector, right-Livingston detector. }
   \label{firstGW}
\end{figure*}

The output of the wavescan transform is a high-resolution wavelet representation in which each sampled time–frequency location (pixel) is characterized by two wavelet amplitudes obtained from the convolution of the time-domain data with the corresponding STFT wavelets (Equations~\ref{FGT1}-\ref{FGT2}). The resolution of the STFT wavelets varies across pixels according to the wavescan algorithm. By adaptively adjusting the resolution, the wavescan transform suppresses artifacts arising from temporal and spectral leakage and yields a high-resolution spectrogram of the time-varying signal power. As an example, Figure~\ref{firstGW}(left) shows the spectrogram of the first GW signal GW150914 detected by the LIGO Hanford and Livingston detectors in 2015~\cite{GW150914}. The characteristic chirping signal from a binary black hole merger is clearly visible in both detectors. The improved time-frequency localization and reduced leakage are particularly advantageous for the identification and reconstruction of GW signals discussed in the next two sections.

\section{Identification of transient events}
\label{cluster}

This section describes the identification of a transient GW signal in data from a network of GW detectors. After the data are conditioned to remove predictable spectral artifacts like quasi-monochromatic lines~\cite{Tiwari2015}, the detector noise is a superposition of quasi-stationary colored noise and transient noise artifacts (glitches) due to the environmental and instrumental disturbances~\cite{LIGOnoise2016}. The power spectral density of the quasi-stationary noise, $|S_k(f)|^2$, is estimated in the WDM domain~\cite{Necula2012} for each detector $k=1,..,K$ in the network. The noise-scaled (whitened) time-series $x_k(t)$ are obtained with the inverse WDM transform. 

A search for a transient GW signal is performed assuming that the characteristic signal waveforms in each detector are not known a priori~\cite{Klimenko2016}. In this case the construction of matched filters is not possible. Therefore, the analysis is performed in the wavescan domain, searching for transient signals whose amplitudes are inconsistent with the fluctuations of the quasi-stationary detector noise. Such wavescan amplitudes can be identified by using the excess-power and cross-power statistics described below.

For each detector $k$, the wavescan power is maximized over all possible time-of-flight delays $\tau_k$ of the anticipated signal, $\tilde{P}_k=\max_{\tau_k}P_k(\tau_k)$. This is used to compute the excess-power for each detector pair $i$ and $j$: 
\begin{align} 
P_{ij} &= r_{i}P_i+r_{j}\tilde{P}_j,~~~ P_i \ge P_j , \label{eq:rk1} \\ 
P_{ij} &= r_{i}\tilde{P}_i+r_{j}P_j,~~~ P_i < P_j , \label{eq:rk2} \\ 
r_{k} &= (1/|S_k|^2)/\sum_i{1/|S_i|^2} , \label{eq:rk3}
\end{align}
 where the weights $r_k$ account for differences in the detector spectral sensitivities $S_k$. The network excess-power is then obtained by summing over all detector pairs, $P_{net}=\sum_{i<j}P_{ij}$. 

To distinguish signals from noise, the cross-power statistic uses the coincident excess power of the GW signals that produce near-simultaneous responses with consistent waveforms in different detectors. The cross-power between the two detectors $i$ and $j$ is defined as
\begin{align} 
C_{ij}= \sqrt{r_ir_jP_i\tilde{P}_j},~~~ P_i \ge P_j , \label{eq:ck1} \\ 
C_{ij}= \sqrt{r_ir_j\tilde{P}_iP_j},~~~ P_i < P_j , \label{eq:ck2}
\end{align}
and, respectively, the network cross-power is defined as
$C_{net}=\sum_{i\neq{j}}C_{ij}$.

\begin{figure*}[t]
   \includegraphics[width=0.49\textwidth]{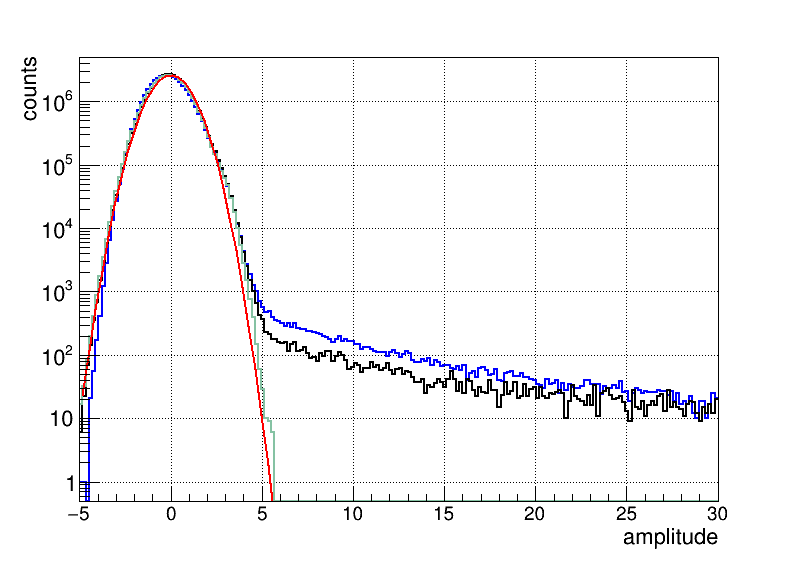}
   \includegraphics[width=0.49\textwidth]{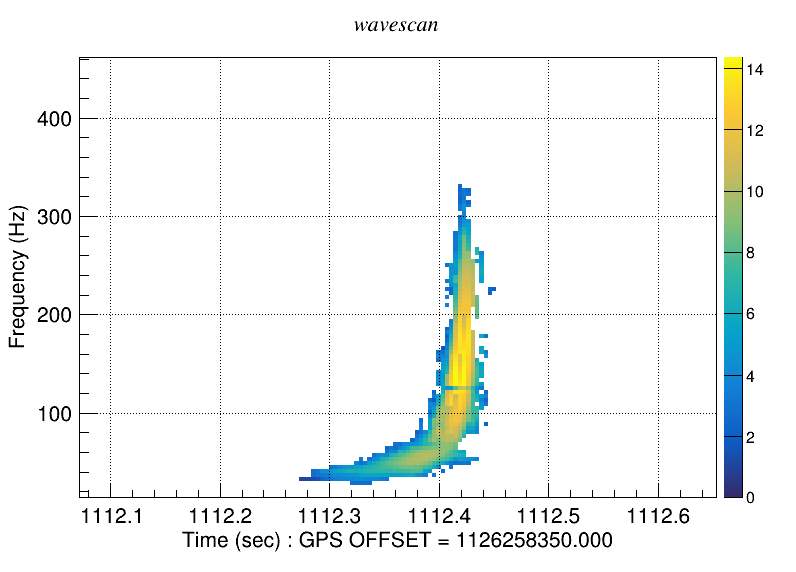}
   \caption{Network-independent statistics for the Hanford-Livingston detector noise. Left: excess-power amplitude $a_p$, black - wavescan, blue - maximum power, teal - wavescan of Gaussian noise, red - $\mathcal{N}(0,1)$ fit. Right: the GW150914 event identified in the wavescan domain by selecting pixels with $a_{pc}>2$ . The color axis shows the cross-power $a_c$ of selected pixels.}
   \label{fig4}
\end{figure*}

For quasi-stationary detector noise, the statistics $P_{net}$, $C_{net}$, and $P_{net}+C_{net}$ approximately follow a Gamma distribution, whose parameters  depend on the number of detectors used in the analysis and their relative sensitivities. To obtain a more universal, network-independent statistic, the following procedure is applied, using $P_{net}$ as an example. First, the Gamma shape parameter $\kappa$ is estimated by the maximum likelihood method~\cite{Gammashape1969}. To reduce glitch-induced bias in $\kappa$, $P_{net}$ samples are capped at a value corresponding to a chosen upper percentile of the distribution. The sample median $\nu_p$ is calculated from the $P_{net}$ distribution. 
\begin{equation}
\label{ap}
a_p^2={({2}\kappa-5/9)\left[(y-1)-log\left( y \right)\right]},~~~~~y=\frac{P_{net}}{\nu_p} \;.
\end{equation} 
The statistic $a_p(y)$ is positive for $y>1$ and it is assigned a negative value for $y<1$. For quasi-stationary detector noise, the $a_p$ amplitudes approximately follow the standard normal distribution $\mathcal{N}(0,1)$ (see Figure~\ref{fig4}). Following the same procedure, the corresponding statistic $a_c$ and $a_{pc}$ can also be constructed for the $C_{net}$ and $P_{net}+C_{net}$ distributions respectively.

Time series from gravitational-wave detectors are highly non-stationary, containing noise transients caused by environmental and instrumental disturbances~\cite{LIGOnoise2016,Davis_2021}. The left panel of Figure~\ref{fig4} shows an example of the $a_p$ distribution for the real noise of the Livingston and Hanford detectors~\cite{AdLIGO2015}. As expected, the excess-power distribution exhibits a Gaussian peak and a long tail from glitches. 
 
Transient events can be identified in the wavescan time-frequency data by excluding pixels in the Gaussian peak and clustering the remaining pixels. Clustered excess-power $\sum_i{a_p^2[i]}$ and cross-power $\sum_i{a_c^2[i]}$ are used for selection of transient events~\cite{2025PhRvD.111b3054M}.
Right panel of Figure~\ref{fig4} shows the cross-power amplitudes for the GW150914 event extracted from the wavescan data. 

The wavescan transform has replaced the WDM transform in the cWB pipeline and has been used in GW searches~\cite{Abac_2025,Abac_2026,wjdz-jdby}. Its computational cost depends on the number of STFT resolutions (typically 6--9) and the STFT oversampling factor (typically 32--64). Although wavescan processes approximately 5--10 times as much time-frequency data as WDM, the overall cWB runtimes are similar for the search configurations considered. A detailed discussion of these searches and their computational performance is beyond the scope of this paper. These searches rely on reconstructing time-domain signals from wavescan data using the inverse wavescan transform, which is described in the next section.

\section{Inverse transform}
\label{waveform}

For a general windowed STFT, the inverse transform can be defined as 
\begin{equation}
   \label{GE}
   {x}(t) = \sum_{n,m}{g_{nm} w_s(t-t_n) e^{i2\pi{f_m}t}} \;, 
\end{equation}
where $w_s(t)$ is the synthesis window, and $g_{nm}$ are the expansion coefficients calculated at each time-frequency location, indexed by 
$n$ and $m$. In the special case of the Gabor expansion, where $w_s(t)$ is a Gaussian window, the coefficients $g_{nm}$ can be calculated by using Equation~\ref{stft} with a suitably chosen analysis window $w_a(t)$~\cite{Bastiaans1980,Wexler1990, Qian1993DiscreteGT}. 
At critical sampling, $\Delta{t}\Delta{f}=1$, a unique window $w_a(t)$ exists such that $g_{nm}$ are given by Equation~\ref{stft}. In general, for an oversampled Gabor transform, $\Delta{t}\Delta{f}<{1}$, the analysis window is no longer unique~\cite{Bastiaans1980}. Several methods have been proposed for calculating the analysis window in the oversampled case~\cite{Bastiaans1980,Wexler1990, Qian1993DiscreteGT}. However, none of these methods apply to the wavescan transform, since its analysis window is defined a priori and the wavescan amplitudes are obtained using wavelets of different durations. Therefore, we seek an approximate solution for the expansion coefficients assuming $w_a(t)={w_s(t)}$, which is expected to hold for highly oversampled STFT transforms~\cite{Wexler1990}. 

A wavescan wavelet is defined to represent the local time-frequency components of the data at time $n$, frequency $m$, and resolution $l$
 \begin{equation}
\label{wsw2}
   {\chi}_{lnm}(t) =  {a_{lnm}\psi_{lnm}(t)} + {\tilde{a}_{lnm}\tilde{\psi}_{lnm}(t)} \;.
\end{equation}
Since only one wavelet is selected from each stack, $l$ is the internal wavescan parameter and omitted below.
The expansion of the time series ${x}(t)$ can be approximated as a superposition of the selected wavescan wavelets  $\chi_{nm}(t)$ 
\begin{equation}
\label{xprime}
   {x}'(t) = \sum_{n,m}\alpha_{nm}\chi_{nm}(t) \;,
\end{equation}
where the $\alpha_{nm}$ are unknown expansion coefficients.
For a Gabor transform described in~\cite{Bastiaans1980,Wexler1990, Qian1993DiscreteGT}, the coefficients $\alpha_{nm}$ are not explicitly calculated, but they are implicitly defined by the calculation of the analysis window. Since this is not possible for the wavescan transform, a different approach is used for the calculation of 
$\alpha_{nm}$. 

\begin{figure}[t]
   \includegraphics[width=0.49\textwidth]{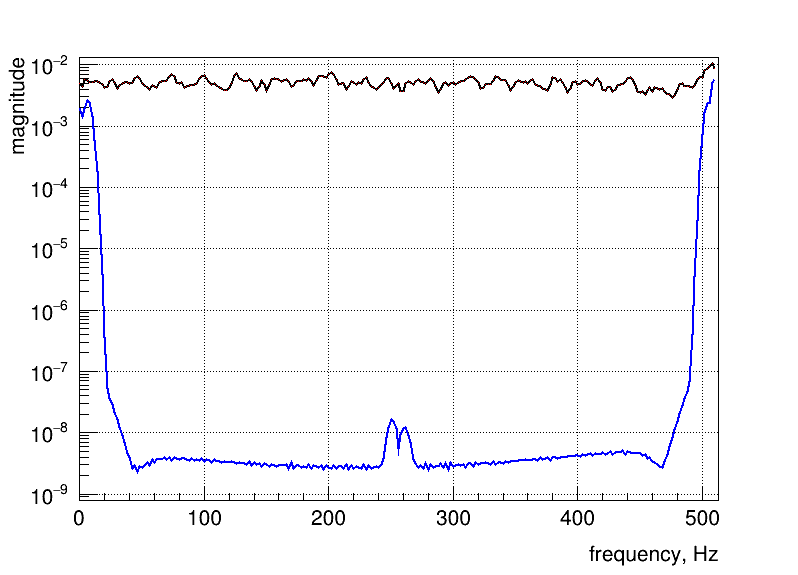}
   \caption{Reconstruction errors vs frequency for the forward-inverse transform of white Gaussian noise. Blue - single-resolution STFT ($T_w=1/8$~s, ${\Delta{t}\Delta{\omega}}=1/32$).  Red - reconstruction errors of the inverse wavescan transform with L=6 resolutions. The residual error is around $16\%$ and is uniform across the frequency band. In both cases the $\alpha^0_{nm}$ expansion coefficients are used.}
   \label{fig6}
\end{figure}

For a highly oversampled transform, one may expect that, to zeroth-order,  the coefficients $\alpha_{nm}$ are approximately equal to 
\begin{equation}
\label{zero}
   \alpha^0_{nm}=\frac{\Delta{t}\Delta{f}}{2} \;.
\end{equation}
\noindent
 For a single resolution STFT with sufficient oversampling, ${\Delta{t}\Delta{f}}<1/16$, the coefficients $\alpha^0_{nm}$ provide a fairly accurate approximation for ${x}'(t)$.  
 For example, applying a forward transform followed by an inverse transform to white Gaussian noise yields a residual ${x}(t)-{x}'(t)$ which approximately follows the normal distribution with the standard deviation $\sigma_0$ of only a few percent (see Table~\ref{tab:1}). 
 As shown in Figure~\ref{fig6}, frequency analysis of the residual indicates that these errors originate at the sampling lattice boundaries. For data, well within the frequency band, the reconstruction errors remain below $10^{-6}$, demonstrating that the expansion coefficients $\alpha^0_{nm}$ enable an accurate inverse transform.
 The same expansion for the wavescan data produces significantly larger errors that are uniform across the frequency band.

To explain these results, note that, unlike an orthogonal transform, Equation~\ref{xprime} represents a sum of strongly correlated wavelets. For a single-resolution STFT, the amplitudes of neighboring wavelets within the regular lattice are precisely balanced to enable accurate reconstruction. In contrast, the wavescan wavelets are selected from a multiresolution set and lose their correlated partners, thus breaking the balance and reducing the reconstruction accuracy. A similar loss occurs for STFT wavelets near the data boundaries, where the expansion coefficients $\alpha^0_{nm}$ are no longer optimal. 
 
This observation indicates that oversampled wavelet data do not admit a unique inverse transform, particularly once the data have been modified by filtering operations. For example, the inverse transform of a selected time-frequency region — such as the region around the GW150914 event shown in Figure~\ref{fig4} (right) — requires different expansion coefficients depending on the size and shape of the selected region, and these coefficients cannot be determined a priori.  Consequently, the expansion coefficients must be recalculated whenever the wavelet data are selected or modified by filtering.
 
The zero-order approximation residual errors ($\sigma_0\approx{16\%}$) are excessive for the wavescan transform where the wavelet duration varies across pixels. It requires a more accurate approximation for $\alpha_{nm}$ which can be found from the minimization of the residual error $|{x}(t)-{x}'(t,\alpha_{nm})|^2$. The resulting system of linear equations for $\alpha_{nm}$ is difficult to solve and, moreover, it does not have a unique solution
\begin{eqnarray}
\label{alpha1}
   {\sum_{\mu\nu}{\alpha_{\mu\nu}\langle{{\chi}_{nm},{\chi}_{\mu\nu}}\rangle}}=\langle{{\chi}_{nm},{\chi}_{nm}}\rangle \;.
\end{eqnarray}
This observation motivates an iterative procedure that converges to a small residual error.
The iterations are based on the cross-talk integrals $\langle{{\chi}_{nm},{\chi}_{\mu\nu}}\rangle$ between a wavelet ${\chi}_{nm}$ and the nearby wavelets.
The cross-talk integrals can be evaluated by using the known overlap of the corresponding STFT wavelets. Higher order approximations $\alpha^i_{nm}$ are obtained by iterations starting from the zero-order expansion coefficients ($i=0$) 
\begin{eqnarray}
\label{alpha0}
   \alpha^{i+1}_{nm}=\alpha^i_{nm} R^i_{nm} \;, ~~R^i_{nm}=\frac{\langle{{\chi}_{nm},{\chi}_{nm}}\rangle}{\sum_{\mu\nu}{\alpha^i_{\mu\nu}\langle{{\chi}_{nm},{\chi}_{\mu\nu}}\rangle}} ,~~
\end{eqnarray}
where the summation is performed over all nearby pixels that have a non-zero cross-talk integral with the pixel $nm$. The iteration procedure is regularized by constraining $R^i_{nm}<4$ and setting the expansion coefficients $\alpha^{i+1}_{nm}=\alpha^0_{nm}$ if $R^i_{nm} \le 0$.

\begin{table}[bht]
    \centering
    \setlength{\tabcolsep}{3pt}
    \begin{tabular}{|c|c|cccccc|}
        \hline
        & &  \multicolumn{6}{c|}{approximations} \\ 
        $l$ & $\Delta{t}\Delta{f}$ & $\sigma_0$ & $\sigma_1$ & $\sigma_2$ & $\sigma_3$  & $\sigma_4$ & $\sigma_5$ \\
        \hline
           4 & 1/32 & $2.1\%$ & $1.2\%$ & $0.9\%$ & $0.7\%$ & $0.6\%$ & $0.4\%$ \\ \hline
        1-6 & 1/16 & $20\%$ & $8.0\%$ & $4.8\%$ & $3.3\%$ & $2.5\%$ & $2.0\%$ \\
        1-6 & 1/32 & $16\%$ & $5.3\%$ & $2.9\%$ & $1.9\%$ & $1.3\%$ & $1.1\%$ \\
        1-7 & 1/32 & $21\%$ & $7.2\%$ & $4.0\%$ & $2.7\%$ & $2.0\%$ & $1.6\%$ \\
        \hline
    \end{tabular}
    \caption{Accuracy of the forward-inverse transform for the single-resolution STFT ($l=4$) and the wavescan transform for different oversampling factors and range of resolutions.  }
    \label{tab:1}
\end{table}

As shown in Table~\ref{tab:1}, for the Gaussian noise test with the single-resolution STFT transform, the consecutive approximations significantly reduce the boundary artifacts reaching the accuracy below $1\%$, and preserve accurate reconstruction of signals that are away from the boundaries.  
This simple and accurate approximation of the inverse STFT transform is sufficient for most signal processing applications that are using an oversampled STFT regression. 

\begin{figure}[t]
   \includegraphics[width=0.49\textwidth]{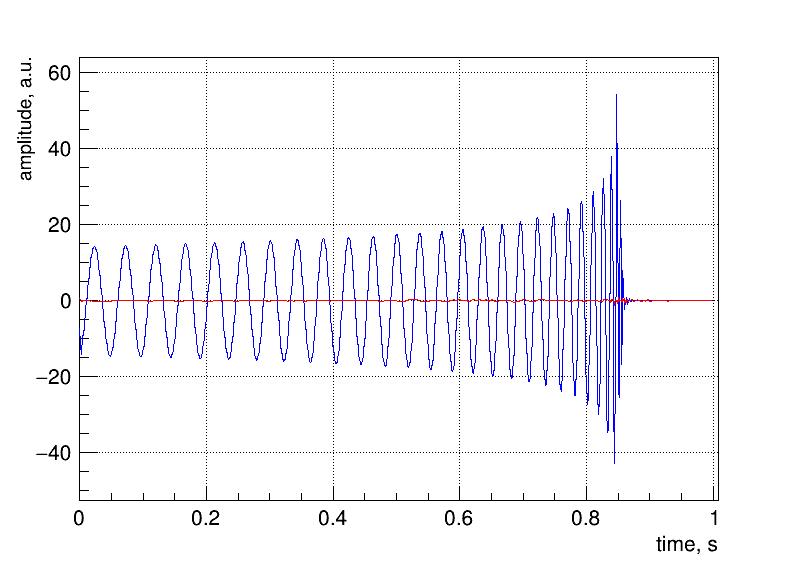}
   \caption{Reconstruction of simulated BBH signal (blue) and the corresponding residual error (red)}
   \label{fig7}
\end{figure}

Higher-order approximations provide reasonable reconstruction accuracy for the inverse wavescan transform as well. The achievable accuracy depends on the wavescan transform parameters. However, for high-order approximations, reconstruction errors can be reduced to a few percent. Reconstruction errors are smaller for deterministic signals, approaching sub-percent accuracy, as demonstrated in Figure~\ref{fig7}, which shows the reconstruction of a simulated BBH signal. While a more accurate inverse wavescan transform can be obtained with additional iterations, the first few iterations are usually sufficient for analysis of GW signals, where reconstruction accuracy is typically limited by uncertainties due to detector noise.

\section{Conclusion}

This paper presents wavescan, an adaptive multiresolution time-frequency transform for the analysis of transient signals in non-stationary data. Wavescan provides a regression framework for gravitational-wave data that includes estimation of the time-dependent signal power, extraction of signals from noise in the time-frequency domain, and reconstruction of their time-domain waveforms. At each time-frequency location, wavescan probes the local signal power using a stack of STFT wavelets with different resolutions and selects the wavelet least affected by temporal and spectral leakage. This adaptive selection produces a high-resolution time-frequency representation with substantially reduced leakage artifacts.
The wavescan representation retains two quadrature amplitudes for every selected time-frequency pixel and therefore contains the phase information required for signal reconstruction.

Two new statistics, based on the excess power and cross-power of multiple detector channels, are introduced. By exploiting the high-resolution localization of transient power in the wavescan domain, these statistics identify time-frequency pixels associated with transient signals and measure the consistency of their responses across the detector network. 

The wavescan time-frequency distribution retains the information required to reconstruct signals in the time domain. A simple and accurate approximation to the inverse STFT is introduced. Based on an iterative reconstruction method, the inverse transform does not depend on a particular choice of window function and can be applied to sufficiently oversampled STFTs which form a stable representation. 
The inverse wavescan transform is less accurate because the wavelet resolution varies across the time-frequency plane. Nevertheless, the tests presented here show that the iterative method reduces the reconstruction error for Gaussian noise to approximately \(1\%-2\%\), depending on the transform parameters, while achieving sub-percent accuracy for the simulated BBH waveform.

These results indicate that wavescan can support the identification, filtering, and approximate reconstruction of transient gravitational-wave signals. Although developed for gravitational-wave data analysis, the method may also be useful for other applications involving identification of transient signals embedded in non-stationary noise.

\section{Acknowledgments}

This research has made use of data, software, and/or web tools obtained from the Gravitational Wave Open Science Center, a service of LIGO Laboratory, the LIGO Scientific Collaboration, and the Virgo Collaboration.
This research was supported by the US National Science Foundation grants PHY-0244902 and PHY-2409372 to the University of Florida, Gainesville, Florida.

\bibliography{references.bib}

@string{prd = {Phys. Rev. D}}

@article{ROBINET2020100620,
title = {Omicron: A tool to characterize transient noise in gravitational-wave detectors},
journal = {SoftwareX},
volume = {12},
pages = {100620},
year = {2020},
issn = {2352-7110},
doi = {https://doi.org/10.1016/j.softx.2020.100620}
}

@article{Qscan2004,
   title={Multiresolution techniques for the detection of gravitational-wave bursts},
   volume={21},
   ISSN={1361-6382},
   url={http://dx.doi.org/10.1088/0264-9381/21/20/024},
   DOI={10.1088/0264-9381/21/20/024},
   number={20},
   journal={Classical and Quantum Gravity},
   publisher={IOP Publishing},
   author={Chatterji, S and Blackburn, L and Martin, G and Katsavounidis, E},
   year={2004},
   month={Sep},
   pages={S1809--S1818}
}

@book{Mallat2008,
author = {Mallat, St{\'e}phane},
title = {A Wavelet Tour of Signal Processing, Third Edition: The Sparse Way},
year = {2008},
isbn = {0123743702},
publisher = {Academic Press, Inc.},
address = {USA},
edition = {3rd},
}

@article{Gabor1946,
	title = {Theory of communications},
	author = {D. Gabor},
	journal = {Journal of the IEE},
	volume = {93},
	pages = {429-457},
	numpages = {28},
	year = {1946},
}

@article{AllenRabiner1977,
  author  = {J. B. Allen and L. R. Rabiner},
  title   = {A Unified Approach to Short-Time Fourier Analysis and Synthesis},
  journal = {Proceedings of the IEEE},
  volume  = {65},
  number  = {11},
  pages   = {1558--1564},
  year    = {1977},
  doi     = {10.1109/PROC.1977.10770}
}

@ARTICLE{Cohen1989,
  author={Cohen, L.},
  journal={Proceedings of the IEEE}, 
  title={Time-frequency distributions review}, 
  year={1989},
  volume={77},
  number={7},
  pages={941-981},
  doi={10.1109/5.30749}
  }

@article{Wigner1932,
	title = {On the quantum correction for thermodynamic equilibrium},
	author = {E. Wigner},
	journal = {Phys. Rev.},
	volume = {40},
	pages = {749-759},
	numpages = {11},
	year = {1932},
}

@article{Ville1948,
	title = {Th{\'e}orie et applications de la notion de signal analytique},
	author = {J. Ville},
	journal = {C{\^a}bles et Transmissions},
	volume = {2A},
	pages = {61-74},
	numpages = {13},
	year = {1948},
}

@article{Cohen1966,
author = {Cohen,Leon },
title = {Generalized Phase-Space Distribution Functions},
journal = {Journal of Mathematical Physics},
volume = {7},
number = {5},
pages = {781-786},
year = {1966},
doi = {10.1063/1.1931206},
url = {https://doi.org/10.1063/1.1931206}
}

@article{Necula2012,
	author  = {Necula, V. and Klimenko, S. and Mitselmakher, G.},
    journal = {Journal of Physics: Conference Series},
	title = {Transient analysis with fast Wilson-Daubechies time-frequency transform},
	volume = {363},
	pages = {012032},
	numpages = {11},
	year = {2012},
	doi = {10.1088/1742-6596/363/1/012032},
}

@article{AdLIGO2015,
	doi = {10.1088/0264-9381/32/7/074001},
	url = {https://doi.org/10.1088/0264-9381/32/7/074001},
    journal = {Classical and Quantum Gravity},
	year = 2015,
	month = {mar},
	publisher = {{IOP} Publishing},
	volume = {32},
	number = {7},
	pages = {074001},
	author = {J Aasi and others},
	collaboration = {LIGO Scientific Collaboration},
	title = {Advanced {LIGO}}
}

@article{AdVirgo2014,
	doi = {10.1088/0264-9381/32/2/024001},
	url = {https://doi.org/10.1088/0264-9381/32/2/024001},
	journal = {Classical and Quantum Gravity},
	year = 2014,
	month = {dec},
	publisher = {{IOP} Publishing},
	volume = {32},
	number = {2},
	pages = {024001},
	author = {F. Acernese and others },
	collaboration = {Virgo Collaboration},
	title = {Advanced Virgo: a second-generation interferometric gravitational wave detector}
}

@article{Gammashape1969,
author = { S. C.   Choi  and  R.   Wette },
title = {Maximum Likelihood Estimation of the Parameters of the Gamma Distribution and Their Bias},
journal = {Technometrics},
volume = {11},
number = {4},
pages = {683-690},
year  = {1969},
publisher = {Taylor & Francis},
doi = {10.1080/00401706.1969.10490731},
}

@ARTICLE{Bastiaans1980,
  author={Bastiaans, M.J.},
  journal={Proceedings of the IEEE}, 
  title={Gabor's expansion of a signal into Gaussian elementary signals}, 
  year={1980},
  volume={68},
  number={4},
  pages={538-539},
  doi={10.1109/PROC.1980.11686}}

@article{Wexler1990,
	title = {Discrete Gabor expansions},
	author = {J. Wexler and S. Raz},
	journal = {Signal Processing},
	volume = {21},
	  number={3},
	pages = {207-221},
	numpages = {14},
	year = {1990},
	month = nov,
}

@ARTICLE{Qian1993DiscreteGT,
  author={Qian, S. and Chen, D.},
  journal={IEEE Transactions on Signal Processing}, 
  title={Discrete Gabor transform}, 
  year={1993},
  volume={41},
  number={7},
  pages={2429-2438},
  doi={10.1109/78.224251}}

@inproceedings {superres,
	title = {A super-resolution spectrogram using coupled {PLCA}},
	booktitle = {INTERSPEECH},
	year = {2010},
	pages = {1696-1699},
	address = {Makuhari, Japan},
	author = {Juhan Nam and Gautham J. Mysore and Joachim Ganseman and Kyogu Lee and Jonathan S. Abel}
}

@ARTICLE{MulSpect1994,
  author={Loughlin, P. and Pitton, J. and Hannaford, B.},
  journal={IEEE Signal Processing Letters}, 
  title={Approximating time-frequency density functions via optimal combinations of spectrograms}, 
  year={1994},
  volume={1},
  number={12},
  pages={199-202},
  doi={10.1109/97.338752}}

@article{superlets,
  title={Time-frequency super-resolution with superlets},
  author={V.~V.~Moca and H.~Bârzan and A.~Nagy-Dabacan and R.~C.~Mureșan},
  journal={Nature Communications},
  year={2021},
  volume={12},
  number = {1},
  pages={337},
  doi = {10.1038/s41467-020-20539-9}
}

@article{GWTC1,
  title = {GWTC-1: A Gravitational-Wave Transient Catalog of Compact Binary Mergers Observed by LIGO and Virgo during the First and Second Observing Runs},
  author = {Abbott, B. P. and others},
  collaboration = {LIGO Scientific Collaboration and Virgo Collaboration},
  journal = {Phys. Rev. X},
  volume = {9},
  issue = {3},
  pages = {031040},
  numpages = {49},
  year = {2019},
  month = {Sep},
  publisher = {American Physical Society},
  doi = {10.1103/PhysRevX.9.031040},
  url = {https://link.aps.org/doi/10.1103/PhysRevX.9.031040}
}

@article{GWTC2,
  title = {GWTC-2: Compact Binary Coalescences Observed by LIGO and Virgo during the First Half of the Third Observing Run},
  author = {Abbott, R and others},
  collaboration = {LIGO Scientific Collaboration and Virgo Collaboration},
  journal = {Phys. Rev. X},
  volume = {11},
  issue = {2},
  pages = {021053},
  numpages = {52},
  year = {2021},
  month = {Jun},
  publisher = {American Physical Society},
  doi = {10.1103/PhysRevX.11.021053},
  url = {https://link.aps.org/doi/10.1103/PhysRevX.11.021053}
}

@article{GWTC3,
  title = {GWTC-3: Compact Binary Coalescences Observed by LIGO and Virgo during the Second Part of the Third Observing Run},
  author = {Abbott, R. and others},
  collaboration = {LIGO Scientific Collaboration, Virgo Collaboration, and KAGRA Collaboration},
  journal = {Phys. Rev. X},
  volume = {13},
  issue = {4},
  pages = {041039},
  numpages = {89},
  year = {2023},
  month = {Dec},
  publisher = {American Physical Society},
  doi = {10.1103/PhysRevX.13.041039},
  url = {https://link.aps.org/doi/10.1103/PhysRevX.13.041039}
}

@ARTICLE{GeomMean,
  author={Cheung, S. and Lim, J.S.},
  journal={IEEE Transactions on Signal Processing}, 
  title={Combined multiresolution (wide-band/narrow-band) spectrogram}, 
  year={1992},
  volume={40},
  number={4},
  pages={975-977},
  doi={10.1109/78.127970}}

@article{Klimenko2005,
  title = {Constraint likelihood analysis for a network of gravitational wave detectors},
  author = {Klimenko, S. and Mohanty, S. and Rakhmanov, M. and Mitselmakher, G.},
  journal = {Phys. Rev. D},
  volume = {72},
  issue = {12},
  pages = {122002},
  numpages = {13},
  year = {2005},
  month = {Dec},
  publisher = {American Physical Society},
  doi = {10.1103/PhysRevD.72.122002},
  url = {https://link.aps.org/doi/10.1103/PhysRevD.72.122002}
}

@article{Klimenko2008,
	doi = {10.1088/0264-9381/25/11/114029},
	url = {https://doi.org/10.1088/0264-9381/25/11/114029},
	journal = {Classical and Quantum Gravity},
	year = 2008,
	month = {may},
	publisher = {{IOP} Publishing},
	volume = {25},
	number = {11},
	pages = {114029},
	author = {S Klimenko and I Yakushin and A Mercer and G Mitselmakher},
	title = {A coherent method for detection of gravitational wave bursts}
	}

@article{Klimenko2016,
  title = {Method for detection and reconstruction of gravitational wave transients with networks of advanced detectors},
  author = {Klimenko, S. and Vedovato, G. and Drago, M. and Salemi, F. and Tiwari, V. and Prodi, G. A. and Lazzaro, C. and Ackley, K. and Tiwari, S. and Da Silva, C. F. and Mitselmakher, G.},
  journal = {Phys. Rev. D},
  volume = {93},
  issue = {4},
  pages = {042004},
  numpages = {10},
  year = {2016},
  month = {Feb},
  publisher = {American Physical Society},
  doi = {10.1103/PhysRevD.93.042004},
  url = {https://link.aps.org/doi/10.1103/PhysRevD.93.042004}
}

@ARTICLE{2025PhRvD.111b3054M,
       author = {{Mishra}, T. and {Bhaumik}, S. and {Gayathri}, V. and {Szczepa{\'n}czyk}, Marek J. and {Bartos}, I. and {Klimenko}, S.},
        title = "{Gravitational waves detected by a burst search in LIGO/Virgo's third observing run}",
      journal = {\prd},
         year = 2025,
        month = jan,
       volume = {111},
       number = {2},
          eid = {023054},
        pages = {023054},
          doi = {10.1103/PhysRevD.111.023054},
archivePrefix = {arXiv},
       eprint = {2410.15191},
 primaryClass = {astro-ph.HE},
       adsurl = {https://ui.adsabs.harvard.edu/abs/2025PhRvD.111b3054M}
}

@article{Abac_2026,
doi = {10.3847/2041-8213/ae447b},
url = {https://doi.org/10.3847/2041-8213/ae447b},
year = {2026},
month = {jun},
publisher = {The American Astronomical Society},
volume = {1004},
number = {2},
pages = {L21},
author = {Abac, A. G.  and others},
collaboration = {LIGO Scientific Collaboration, Virgo Collaboration, and KAGRA Collaboration},
itle = {GWTC-4.0: Methods for Identifying and Characterizing Gravitational-wave Transients},
journal = {The Astrophysical Journal Letters}
}

@article{wjdz-jdby,
  title = {All-sky search for short gravitational-wave bursts in the first part of the fourth LIGO-Virgo-KAGRA observing run},
  author = {Abac, A. G. and others},
  collaboration = {LIGO Scientific Collaboration, Virgo Collaboration, and KAGRA Collaboration},
  journal = {Phys. Rev. D},
  volume = {112},
  issue = {10},
  pages = {102005},
  numpages = {29},
  year = {2025},
  month = {Nov},
  publisher = {American Physical Society},
  doi = {10.1103/wjdz-jdby},
  url = {https://link.aps.org/doi/10.1103/wjdz-jdby}
}

@article{Abac_2025,
doi = {10.3847/1538-4357/adc681},
url = {https://doi.org/10.3847/1538-4357/adc681},
year = {2025},
month = {may},
publisher = {The American Astronomical Society},
volume = {985},
number = {2},
pages = {183},
author = {Abac, A. G. and others},
collaboration = {LIGO Scientific Collaboration, Virgo Collaboration, and KAGRA Collaboration},
title = {Search for Gravitational Waves Emitted from SN 2023ixf},
journal = {The Astrophysical Journal}
}

@article{GW150914,
  title = {Observation of Gravitational Waves from a Binary Black Hole Merger},
  author = {Abbott, B. P. and others},
  collaboration = {LIGO Scientific Collaboration and Virgo Collaboration},
  journal = {Phys. Rev. Lett.},
  volume = {116},
  issue = {6},
  pages = {061102},
  numpages = {16},
  year = {2016},
  month = {Feb},
  publisher = {American Physical Society},
  doi = {10.1103/PhysRevLett.116.061102},
  url = {https://link.aps.org/doi/10.1103/PhysRevLett.116.061102}
}

@article{GW150914properties,
  title = {Properties of the Binary Black Hole Merger GW150914},
  author = {Abbott, B. P. and others},
  collaboration = {LIGO Scientific Collaboration and Virgo Collaboration},
  journal = {Phys. Rev. Lett.},
  volume = {116},
  issue = {24},
  pages = {241102},
  numpages = {19},
  year = {2016},
  month = {Jun},
  publisher = {American Physical Society},
  doi = {10.1103/PhysRevLett.116.241102},
  url = {https://link.aps.org/doi/10.1103/PhysRevLett.116.241102}
}

@article{LIGOnoise2016,
	doi = {10.1088/0264-9381/33/13/134001},
	url = {https://doi.org/10.1088/0264-9381/33/13/134001},
	journal = {Classical and Quantum Gravity},
	year = 2016,
	month = {jun},
	publisher = {{IOP} Publishing},
	volume = {33},
	number = {13},
	pages = {134001},
	author = {B P Abbott and others},
	collaboration = {LIGO Scientific Collaboration and Virgo Collaboration},
	title = {Characterization of transient noise in Advanced {LIGO} relevant to gravitational wave signal {GW}150914}
}

@article{Davis_2021,
  author  = {Davis, Derek and Areeda, Joseph S. and Berger, Beverly K.
             and Bruntz, Robert and Effler, Anamaria and Essick, Reed C.
             and Fisher, Ryan P. and Godwin, Patrick and Goetz, Evan
             and Helmling-Cornell, Adrian F.},
  title   = {{LIGO} detector characterization in the second and third observing runs},
  journal = {Classical and Quantum Gravity},
  volume  = {38},
  number  = {13},
  pages   = {135014},
  year    = {2021},
  doi     = {10.1088/1361-6382/abfd85},
  url     = {https://doi.org/10.1088/1361-6382/abfd85}
}

@article{Tiwari2015,
	doi = {10.1088/0264-9381/32/16/165014},
	url = {https://doi.org/10.1088/0264-9381/32/16/165014},
	journal = {Classical and Quantum Gravity},
	year = 2015,
	month = {jul},
	publisher = {{IOP} Publishing},
	volume = {32},
	number = {16},
	pages = {165014},
	author = {V Tiwari and M Drago and V Frolov and S Klimenko and G Mitselmakher and V Necula and G Prodi and V Re and F Salemi and G Vedovato and I Yakushin},
	title = {Regression of environmental noise in {LIGO} data}
}

\end{document}